\documentclass{article}
\usepackage[preprint]{spconf}
\copyrightnotice{\copyright\ IEEE 2021}
\toappear{To appear in {\it Proc.\ ICASSP2021,
                    June 06-11, 2021, Toronto, Ontario, Canada}}
\usepackage{url}
\usepackage{amsmath,graphicx}
\usepackage{amssymb}
\usepackage{cite}
\usepackage{subcaption}
\usepackage{graphicx}
\usepackage{hyperref}
\usepackage{cleveref}
\usepackage{mathtools}
\usepackage{pgfplots}
\usepackage{physics}
\usepackage{textcomp}
\usepackage[ruled]{algorithm2e}
\pgfplotsset{compat=newest}
\usetikzlibrary{spy,backgrounds}
\usepgfplotslibrary{units}
\usepackage{pgfplotstable}
\title{Parallel Iterated Extended and Sigma-Point Kalman Smoothers}
\name{Fatemeh Yaghoobi, Adrien Corenflos, Sakira Hassan, Simo Särkkä\thanks{The authors would like to thank Academy of Finland for funding.}}
\address{Department of Electrical Engineering and
Automation, Aalto University, Finland}
\begin{document}
\maketitle 
\begin{abstract}
The problem of Bayesian filtering and smoothing in nonlinear models with additive noise is an active area of research. Classical Taylor series as well as more recent sigma-point based methods are two well-known strategies to deal with these problems. However, these methods are inherently sequential and do not in their standard formulation allow for parallelization in the time domain. In this paper, we present a set of parallel formulas that replace the existing sequential ones in order to achieve lower time (span) complexity. Our experimental results done with a graphics processing unit (GPU) illustrate the efficiency of the proposed methods over their sequential counterparts.
\end{abstract}
\section{Introduction}
\label{sec:intro}
In recent years, the rapid advancements in hardware technologies such as graphics processing units (GPUs) and tensor processing units (TPUs) allow compute-intensive workloads to be offloaded from the central processing units (CPUs) by introducing parallelism \cite{rauber2013parallel, owens2008gpu, jouppi2017datacenter}. There is a wide variety of areas that can benefit from parallelization \cite{cormen2009introduction}, one of which is state estimation.

State estimation is a common task that arises in various areas of science and engineering  \cite{sarkka2013bayesian,Bar-Shalom+Li+Kirubarajan:2001,jazwinski1970stochastic}. It aims at combining the noisy measurements and the model to estimate the hard-to-measure states. A frequent and classical method for solving this problem is based on Bayesian filtering and smoothing \cite{sarkka2013bayesian} which inherently provides a sequential solution with linear complexity in the number of time steps. 

In order to tackle the computational burden of Kalman type of filters and smoothers, \cite{barfoot2014batch,grigorievskiy2017parallelizable} provide sub-linear computational methods by taking advantage of the sparse structures of the matrices appearing in the batch forms of the problems. In other works, using an ensemble formulation of Kalman filter has been used to speed up the matrix computations through parallelization \cite{ghorbanidehno2020recent, evensen2003ensemble}. The primary focus of these works was the efficient computation of the covariance matrices either by introducing sparse or sample covariance matrices rather than considering the temporal state-space structure per se. While in the aforementioned works, parallelization of the sub-problems in the area of Bayesian filtering and smoothing were considered, \cite{sarkka2020temporal} presented a general parallelizable formulations specifically designed for parallelizing state-estimation problems in the temporal direction. Moreover, for the special case of linear Gaussian model, parallel equations for computing Kalman filter and Raugh--Tung--Striebel smoother solutions were derived. 

Overcoming the computational burden in the case of nonlinear dynamical systems with additive Gaussian noise is also of paramount importance. In these types of models, various linearization approaches can be used. Taylor series expansion based iterated extended Kalman smoother (IEKS) methods \cite{bell1993iterated,bell1994iterated,sarkka2020levenberg} and sigma-point based methods \cite{sarkka2013bayesian} are well-established techniques in literature. Iterated sigma-point methods have been proposed, for example, in  \cite{garcia2016iterated, garcia2015posterior}. Despite the capabilities of the aforementioned methods in state estimation in nonlinear Gaussian models, they lack a framework which enables the computations in a more efficient way when using parallelization.

The contribution of this paper is to present a set of parallelizable formulas for filtering and smoothing in nonlinear Gaussian systems, in particular, IEKS and sigma-point based methods using a scan algorithm \cite{blelloch1989scans,sarkka2020temporal}. The proposed methods reduce the linear span complexity of the state estimation methods to logarithmic with respect to the number of measurements.

This paper is organized as follows: Section~\ref{sec:background} briefly reviews the generic parallel framework for Bayesian filters and smoothers. Sections~\ref{sec:problem formulation} and \ref{sec:proposed method} are concerned with presenting the formulation of the problem and proposing our method. Section \ref{sec:Experimental-results} analyzes the efficiency and the computational complexity of the proposed method through one numerical example, and Section~\ref{sec:conclusion} concludes the paper.
\section{General Parallel Framework for Bayesian Filters and Smoothers}
\label{sec:background}
It is shown in \cite{sarkka2020temporal} that the computation of sequential Bayesian filtering and smoothing can be converted to general parallel formulas in terms of associative operations. This allows for the use of the parallel scan method \cite{blelloch1989scans} which is a common algorithm used to speed-up sequential computations, for example, on GPU-based computing systems. In the rest of this section, we review the general parallel algorithms provided in \cite{sarkka2020temporal} which we then extend to nonlinear Gaussian models.

Given a state space model of the following form:
\begin{equation} \label{eq:state-space-model}
  x_k \sim p(x_k \mid x_{k-1}), \quad y_k \sim p(y_k \mid x_k),
\end{equation}
the goal of the filtering problem is to find the posterior distributions $ p(x_k\mid y_{1:k})$ for $k=1,\ldots,n$. This distribution is a probabilistic representation of the available statistical information on the state $x_k \in \mathbb{R}^{n_x}$ given the measurements $y_{1:k} = \{ y_1,\ldots,y_k \}$ with $y_k \in \mathbb{R}^{n_y}$. Having acquired the filtering results for $k=1,\ldots,n$, and using all the $n$ measurements, the Bayesian smoother can be used to compute the posterior distributions $p(x_k\mid y_{1:n})$. The following strategies are used in \cite{sarkka2020temporal} so as to particularize $a_k$ and the binary associative operator $\otimes$ which provide a parallel framework for solving the aforementioned sequential filtering and smoothing problem.
\paragraph*{\textbf{Filtering.}}
Given two positive functions $g_i^ \prime (y), g_j^ \prime(y)$ and two conditional densities $f_i^ \prime (x \mid y), f_j^ \prime (x \mid y)$, the authors of \cite{sarkka2020temporal} proved that the binary operation $(f_i^ \prime,g_i^ \prime) \otimes (f_j^ \prime,g_j^ \prime) = (f_{ij}^ \prime,g_{ij}^ \prime)$ defined by 
\begin{equation}
\begin{split}
f_{ij}^ \prime(x|z)=\frac{\int{g_i^ \prime(y)f_j^ \prime(x \mid y)}f_i^ \prime(y\mid z)dy}{\int{g_j^ \prime(y)}f_i^ \prime(y \mid z)dy},\\
g_{ij}^ \prime(z)=g_i^ \prime(z)\int{g_j^ \prime(y)f_i^ \prime(y \mid z)dy},
\end{split}
\end{equation}
is associative and by selecting $a_k = (f_k^ \prime,g_k^ \prime)$ as follows:
\begin{equation}  \label{a_k_filter}
\begin{split}
  f_k^ \prime(x_k \mid x_{k-1}) &= p(x_k \mid y_k, x_{k-1}), \\
  g_k^ \prime(x_{k-1}) &= p(y_k \mid x_{k-1}),
\end{split}
\end{equation}
where $p(x_1 \mid y_1,x_0) = p(x_1 \mid y_1)$ and $p(y_1 \mid x_0)= p(y_1)$, the Bayesian map $\binom{p(x_k|y_{1:k})}{p(y_{1:k})}$ can be rewritten as the $k$-th prefix sum, $a_1\otimes \dots \otimes a_k$.
\paragraph*{\textbf{Smoothing.}}
Similarly \cite{sarkka2020temporal}, for any conditional densities $f_i^ \prime(x \mid y)$ and $f_j^ \prime(x \mid y)$ the binary operation $f_i^ \prime \otimes f_j^ \prime \coloneqq \int{f_i^ \prime(x|y)f_j^ \prime(y|z)}dy$ is associative and by selecting $a_k = p(x_k \mid y_{1:k}, x_{k+1})$
with $a_n=p(x_n \mid y_{1:n})$, the Bayesian smoothing solution can then be calculated as $p(x_k \mid y_{1:n}) = a_k \otimes a_{k+1} \otimes \dots \otimes a_n $.

Having considered the aforementioned general formulations, in this paper, we aim to extend the element $a_k$ and the binary associative operator $ \otimes $ to linear approximations of non-linear Gaussian systems, specifically, to the extended Kalman filter and smoother, and sigma-points methods.
\section{Problem formulation}
\label{sec:problem formulation}
We consider the following model:
\begin{equation} \label{nonlinear/gaussian model}
\begin{split}
  x_k &= f_{k-1}(x_{k-1})+q_{k-1},\\     
  y_k &= h_k(x_k)+r_k,
\end{split}
\end{equation}
where $f_{k-1}(.)$ and $h_{k}(.)$ are nonlinear functions. The $q_k$ and $r_k$ are the process and measurement noises, which are assumed to be zero-mean, independent Gaussian noises with known covariance matrices, $Q_k$ and $R_k$, respectively. Furthermore, the initial state is Gaussian $x_0 \sim N(m_0 , P_0)$ with known mean $m_0$ and covariance $P_0$. This paper is concerned with the computing approximate posterior distributions of the states $x_{0:n}=\{x_0,x_1,\ldots,x_n\}$ given all the measurements $y_{1:n}=\{y_1,x_1,\ldots,y_n\}$ in parallel form, or more precisely, the corresponding filtering and smoothing distributions.

Since the filtering and smoothing problems are not solvable in closed-form in the general non-linear case, one needs to resort to approximations. Here we follow the Gaussian filtering and smoothing frameworks \cite{sarkka2013bayesian} and form linear approximations of the system \eqref{nonlinear/gaussian model} in the following form:
\begin{equation} \label{eq:enabaling-approximation}
\begin{split}
  f_{k-1}(x_{k-1}) &\approx F_{k-1}x_{k-1} + c_{k-1} + e_{k-1},\\
  h_k(x_k) &\approx H_k x_k + d_k + v_k,
\end{split}
\end{equation}
where $F_k \in \mathbb{R}^{n_x\times n_x}$, $c_k \in \mathbb{R}^{n_x}$, $H_k \in \mathbb{R}^{n_y \times n_x}$, $d_k \in \mathbb{R}^{n_y}$, $e_k \in \mathbb{R}^{n_x}$ and $v_k \in \mathbb{R}^{n_y}$ are zero mean Gaussian noises with covariance matrices $\Lambda_k$ and $\Omega_k$, respectively. 

There are different strategies to effectively select the parameters of \eqref{eq:enabaling-approximation}. In this paper, we will consider two such strategies widely-used in the Gaussian filtering literature, namely iterated sigma-point and extended Kalman smoothers \cite{bell1994iterated,sarkka2020levenberg,garcia2016iterated}. In these approaches, the linearized-filter-smoother method is repeated $M$ times, with the linearization parameters leveraging the results of the previous smoothing pass instead of the previous step. We can therefore see our successive linear approximations as being parametrized by the following vectors and matrices:
\begin{equation}\label{eqn:linear-param}
    F_{0:n-1}^{(i)}, c_{0:n-1}^{(i)}, \Lambda_{0:n-1}^{(i)}, H_{0:n-1}^{(i)}, d_{1:n}^{(i)}, \Omega_{1:n}^{(i)}.
\end{equation}
In the rest of this section, we will discuss how to acquire the linearized parameters of \eqref{eqn:linear-param} using these methods. Also, for the sake of notational simplicity, we drop the index $i$ from these parameters.
\paragraph*{\textbf{Iterated sigma-point method.}}
In this approach, we select the parameters $(F_{k-1},c_{k-1},\Lambda_{k-1})$ and $(H_k,d_k,\Omega_k)$ using sigma-point-based statistical linear regression (SLR) method \cite{garcia2016iterated} as follows. First, we select $m$ sigma points $\mathcal{X}_{1,k}^{(i)},\ldots,\mathcal{X}_{m,k}^{(i)}$ and their associated weights $w_{1,k}^{(i)},\ldots,w_{m,k}^{(i)}$ according to the posterior moments $\bar{x}_k^{(i-1)}$ and $\bar{P}_k^{(i-1)}$ of the previous iteration, which are the best available estimates for the means and covariances of the smoothing distribution. Then, in order to find the parameters $(F_{k-1},c_{k-1},\Lambda_{k-1})$, transformed sigma-points are obtained as $ \mathcal{Z}_j = f_{k-1}(\mathcal{X}_{j,k-1}^{(i)})$ for $j=1, \ldots,m$, and the linearization parameters are then given by:
\begin{equation}\label{eq:propag_param}
 \begin{split}
F_{k-1} &= \Psi^\top {\bar{P}_{k-1}}^{-1}, \\
c_{k-1} &= \bar{z} - F_{k-1} \bar{x}_{k-1},  \\
\Lambda_{k-1} &= \Phi - F_{k-1} \bar{P}_{k-1} F_{k-1}^\top.
 \end{split}
\end{equation}
If we now write $\bar{x}=\bar{x}_{k-1}$ and $w_j = w_{j,k-1}^{(i)}$, the required moment approximations for Equation~\eqref{eq:propag_param} are \cite{arasaratnam2007discrete}:
\begin{equation} \label{eq:SLR-moments}
\begin{split}
\bar{z} &\approx \sum_{j=1}^{m} w_j \mathcal{Z}_j,\\
\Psi &\approx \sum_{j=1}^{m} w_j (\mathcal{X}_j - \bar{x})(\mathcal{Z}_j- \bar{z})^\top,\\
\Phi &\approx \sum_{j=1}^{m} w_j (\mathcal{Z}_j- \bar{z})(\mathcal{Z}_j- \bar{z})^\top.
\end{split}
\end{equation}
Similarly, reusing Equations~\eqref{eq:SLR-moments} with $\bar{x} = \bar{x}_{k}$, $w_j = w_{j,k}^{(i)}$, and $\mathcal{Z}_j = h_k(\mathcal{X}_{j,k}^{(i)})$ the parameters $(H_k,d_k,\Omega_k)$ can be calculated as follows:
\begin{equation}\label{eq:obs_param}
\begin{split}
H_k &= \Psi^\top {\bar{P}_{k}}^{-1}, \\
d_k &= \bar{z} - H_k \bar{x}_k, \\
\Omega_k &= \Phi - H_k \bar{P}_{k} H_k^\top.
\end{split}
\end{equation}
The iterated posterior linearization smoother (IPLS) \cite{garcia2016iterated} now consists in iterating Equations \eqref{eq:propag_param} and \eqref{eq:obs_param} with updated approximate means and covariances of the posterior distribution at each iteration. 
\paragraph*{\textbf{Iterated extended Kalman smoother.}}
In this case, $\Omega$ and $\Lambda$ are selected as zeros, and $(F_{k-1},c_{k-1})$ and $(H_k,d_k)$ are obtained by analytical linearization at the previous posterior (smoother) mean estimate of $x_{0:N}$. This approach is recognized as Gauss--Newton method when computing the MAP estimates \cite{bell1994iterated} and it can also be extended to correspond to Levenberg--Marquardt method \cite{sarkka2020levenberg}. Here, we aim to obtain the linearized parameters according to this method which will be used in the next section to get parallel formulas.

By expanding $f_{k-1}(x_{k-1})$ and $h_k(x_k)$ in the first-order Taylor series utilizing the previous posterior means $\bar{x}_k$, the parameters of \eqref{eqn:linear-param} are:
\begin{equation} \label{eq:IEKF-param}
\begin{split}
F_{k-1} &= \nabla f(\bar{x}_{k-1}), \\
c_k &= f(\bar{x}_{k-1}) - F_{k-1} \bar{x}_{k-1},\\
H_{k} &= \nabla h({\bar{x}_{k}}), \\
d_k &= h({\bar{x}_{k}}) - H_{k} \bar{x}_{k},
\end{split}
\end{equation}
where $\nabla f$ and $\nabla h$ are the Jacobians of $f$ and $h$, respectively. Please note that in this paper computation of parameters in \eqref{eq:propag_param} and \eqref{eq:obs_param}, and \eqref{eq:IEKF-param} is performed offline, which means that we have all measurements as well as the results of previous trajectory, that is, $\bar{x}_{1:n}$ and $\bar{P}_{1:n}$ for all $n$ data points.

Having obtained the linearized parameters, the remaining task is to find the parallel formulas which will be discussed in the next section.
\section{The proposed method}
\label{sec:proposed method}
Probability densities for the model of form \eqref{nonlinear/gaussian model} with linearization parameters of form \eqref{eqn:linear-param} can be formulated as follows:
\begin{equation}  \label{eq:nonlinear-gaussian-model}
\begin{split}
 p(x_k \mid x_{k-1}) &\approx N(x_k; F_{k-1}x_{k-1} + c_{k-1}, Q^\prime_{k-1}), \\
 p(y_k \mid x_k) &\approx N(y_k; H_k x_k+ d_k, R^\prime_k),
 \end{split}
\end{equation}
where $Q_{k-1}^\prime = Q_{k-1} + \Lambda_{k-1}$ and $R_k^\prime = R_k + \Omega_{k}$. The goal here is to obtain the parallel nonlinear Gaussian filter and smoother for the model \eqref{eq:nonlinear-gaussian-model}. To meet this goal, similar to the method used in \cite{sarkka2020temporal}, we define $a_k$ and binary operator $\otimes$ for our new linearized model.
\paragraph*{\textbf{Nonlinear Gaussian filtering.}}
Aiming to specify the element $a_k$ for obtaining parallel filtering equations according to \eqref{a_k_filter}, we apply Kalman filter update step to the density $p(x_k \mid x_{k-1})$ with measurement $y_k$. The results of the matching terms are as follows:
\begin{equation}
    \begin{split}
       f_k^ \prime(x_k \mid x_{k-1}) &= p(x_k \mid  y_k,x_{k-1})\\
       &=N(x_k;A_k x_{k-1} + b_k,C_k),
    \end{split}
\end{equation}
where:
\begin{equation} \label{eq:parallel-parameters-filtering-1}
\begin{split}
A_k &= (I_{n_x} - K_k H_k)F_{k-1}, \\
b_k &= c_{k-1} + K_k( y_k - H_k c_{k-1} - d_{k-1}), \\
C_k & = (I_{n_x} -K_k H_k) Q_{k-1} ^ \prime, \\
K_k &= Q_{k-1} ^ \prime {H_k}^\top S_k^{-1},\\
S_k &= H_k Q_{k-1} ^ \prime H_k^\top+ R_k ^ \prime.
\end{split}
\end{equation}
It is worth noticing that in order to find parameters of \eqref{eq:parallel-parameters-filtering-1} at $k=1$ and given $m_0$ and $P_0$, conventional formulations of the Kalman filter method with the linearized parameters are applied directly for prediction and update steps.

Also, using the information form of Kalman filter \cite{Anderson+Moore:1979}, the distribution $g_k^ \prime(x_{k-1}) = p(y_k \mid x_{k-1})\propto N_I(x_{k-1}; \eta_k,J_k)$ can be obtained as follows:
\begin{equation} \label{eq:parallel-parameters-filtering-2}
\begin{split}
    J_k &= (H_k F_{k-1})^\top S_k^{-1} H_k F_{k-1} \\
    \eta_k &= (H_k F_{k-1} )^\top S_k^{-1} H_k  (y_k - H_k c_{k-1} - d_k).
\end{split}
\end{equation}
Equations~\eqref{eq:parallel-parameters-filtering-1} and \eqref{eq:parallel-parameters-filtering-2} provide the parameters of element $a_k=(A_k,b_k,C_k,\eta_k,J_k)$ in the filtering step, and they can be computed in parallel. Also, given $a_i$ and $a_j$ with the mentioned parameters, the binary associative operator $a_i \otimes a_j = a_{ij}$ can then be calculated with the following parameterization \cite[lemma 8]{sarkka2020temporal}:
\begin{equation} \label{eq:op-filter}
\begin{split}
A_{ij} &= A_j (I_{n_x} + C_i J_j)^{-1} A_i,\\
b_{ij} &= A_j (I_{n_x} + C_i J_j)^{-1} (b_i + C_i \eta_j) + b_j,\\
C_{ij} &= A_j (I_{n_x} + C_i J_j)^{-1} C_i A_j^\top + C_j,\\
\eta_{ij} &= A_i^\top (I_{n_x} + J_j C_i)^{-1} (\eta_j - J_j b_i) + \eta_i,\\
J_{ij} &= A_i^\top (I_{n_x} + J_j C_i)^{-1} J_j A_i + J_i.
\end{split}
\end{equation}
The proof for Equations~\eqref{eq:op-filter}  can be found in \cite{sarkka2020temporal}.
\paragraph*{\textbf{Nonlinear Gaussian smoothing.}}
Assume that the filtering means $x^*_k$ and covariance matrices $P^*_k$ for the model \eqref{eq:nonlinear-gaussian-model} have been acquired as described above. We now get the following parameters for the smoothing step:
\begin{equation}
\begin{split}
p(x_k \mid y_{1:k}, x_{k+1}) &= N(x_k; E_k x_{k+1} + g_k, L_k)\\
\end{split}
\end{equation}
for $k<n$:
\begin{equation}
\begin{split}
E_k &= P_k F_k^\top  (F_k P^*_k F_k^\top + Q_{k-1}^\prime )^{-1}, \\
g_k &= x^*_k - E_k (F_k x^*_k + c_k), \\
L_k &= P^*_k - E_k F_k P^*_k,
\end{split}
\end{equation}
and for $k=n$:
\begin{equation}
\begin{split}
E_n &= 0,\\
g_n &= x^*_n,\\
L_n &= P^*_n.
\end{split}
\end{equation}
In the smoothing step, the parameters $a_k = (E_k,g_k,L_k)$ can be calculated in parallel. Now, given two elements $a_i$ and $a_j$,  the binary associative operator defined by $a_i \otimes a_j = a_{ij}$ can be parametrized  as follows \cite[lemma 10]{sarkka2020temporal}: 
\begin{equation} \label{eq:op-smoother}
\begin{split}
E_{ij} &= E_i E_j,\\
g_{ij} &= E_i g_j + g_i,\\
L_{ij} &= E_i L_j E_i^\top + L_i.
\end{split}
\end{equation}
\section{Experimental results}
\label{sec:Experimental-results}
In this section, we evaluate the performance of the proposed methods on a simulated coordinated turn model with a bearings only measurement model \cite{bar1995multitarget} which was also used in \cite{sarkka2020levenberg}.
To this end, we compare the effective average run time of the parallel versions of the extended (IEKS) and cubature integration \cite{sarkka2013bayesian} based sigma-point iterated smoothers (IPLS) with $M=10$ iterations, as described in Section~\ref{sec:proposed method}, with their sequential counterparts both on a CPU (Intel\textsuperscript{\textregistered} Xeon\textsuperscript{\textregistered} running at 2.30GHz) and on a GPU (Nvidia\textsuperscript{\textregistered} Tesla\textsuperscript{\textregistered} P100 PCIe 16 GB with 3584 cores). For our experiments we leverage the JAX framework \cite{jax2018github} which implements the Blelloch parallel-scan algorithm \cite{blelloch1989scans} natively\footnote{The code to reproduce the experiments can be found at the following \href{https://github.com/EEA-sensors/parallel-non-linear-gaussian-smoothers}{address}.}.

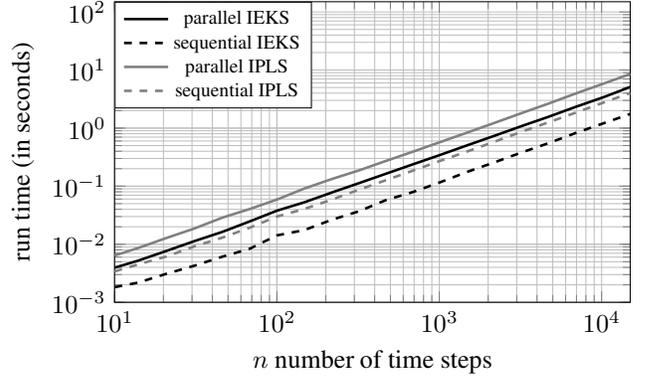
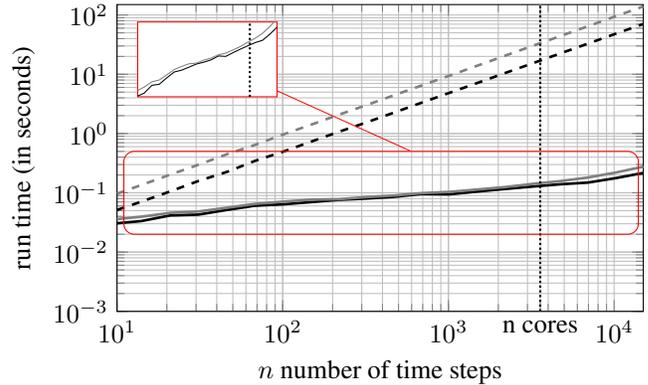
\begin{figure}[!ht]
    \centering
    \begin{subfigure}[b]{\linewidth}
        \centering
        \resizebox{\textwidth}{!}{\begin{tikzpicture}[scale=0.33]
\begin{axis}[
    width=\textwidth,
    height=0.66\textwidth,
    legend style={
        nodes={scale=0.75, transform shape},
        at={(0,1)},
        anchor=north west},
    grid=both,
    xmode=log,
    ymode=log,
    xmin=10, xmax=1.5e4,
    % restrict y to domain*=1E-2:1E0,
    ymin=1e-3, ymax=150,
    xlabel={$n$ number of time steps},
    ylabel={run time (in seconds)},
    ]
\addplot[black, line width=1pt] table [x=time_steps, y=CPU_par_IEKS, col sep=comma]{runtime.csv};
\addplot[black, dashed, line width=1pt] table [x=time_steps, y=CPU_seq_IEKS, col sep=comma]{runtime.csv};
\addplot[gray, line width=1pt] table [x=time_steps, y=CPU_par_ICKS, col sep=comma]{runtime.csv};
\addplot[gray, dashed, line width=1pt] table [x=time_steps, y=CPU_seq_ICKS, col sep=comma]{runtime.csv};
\legend{parallel IEKS, sequential IEKS, parallel IPLS, sequential IPLS};
\end{axis}
\end{tikzpicture}}
        \caption{CPU run time}
        \label{fig:cpu}
    \end{subfigure}\hfill
    \begin{subfigure}[b]{\linewidth}
        \centering
        \resizebox{\textwidth}{!}{\begin{tikzpicture}
% \begin{scope}[spy using outlines={rectangle, magnification=1.25,
%   width=0.5\textwidth,height=0.5\textwidth,connect spies}]
\begin{axis}[
    width=\textwidth,
    height=0.66\textwidth,
    % legend style={
    %     nodes={scale=0.75, transform shape},
    %     at={(1,1)},
    %     anchor=south east},
    grid=both,
    xmode=log, 
    ymode=log,
    xmin=10, xmax=1.5e4,
    % restrict y to domain*=1E-2:1E0,
    ymin=1e-3, ymax=150,
    extra x ticks={3584},
    extra x tick style={%
        grid=major,
    },
    extra x tick labels={n cores},
    xlabel={$n$ number of time steps},
    ylabel={run time (in seconds)},
    ]
\addplot[black, line width=1pt] table [x=time_steps, y=GPU_par_IEKS, col sep=comma]{runtime.csv};
\addplot[black, dashed, line width=1pt] table [x=time_steps, y=GPU_seq_IEKS, col sep=comma]{runtime.csv};
\addplot[gray, line width=1pt] table [x=time_steps, y=GPU_par_ICKS, col sep=comma]{runtime.csv};
\addplot[gray, dashed, line width=1pt] table [x=time_steps, y=GPU_seq_ICKS, col sep=comma]{runtime.csv};
\addplot +[black, thick, densely dotted, mark=none] coordinates {(3584, \pgfkeysvalueof{/pgfplots/ymin}) (3584, \pgfkeysvalueof{/pgfplots/ymax})};
% \legend{parallel IEKS, sequential IEKS, parallel ICKS, sequential ICKS}
\draw[draw=red,rounded corners] (axis cs:11,2E-2) rectangle (axis cs:1.4E4,0.5);
\coordinate (pt) at (axis cs:600,0.5);
\end{axis}
\node[inner sep=0pt,outer sep=0pt, pin={[inner sep=0pt,outer sep=0pt, draw=red, fill=white,anchor=south,pin edge={red}, pin distance=2.8cm]165:{%
    \begin{tikzpicture}[trim axis left,trim axis right]
    \begin{axis}[
        % hide axis,
        % axis line style={red},
        height=0.3\textwidth,
        width=0.4\textwidth,
        grid=both,
        xmode=log, 
        ymode=log,
        ymin=3E-2,ymax=0.25,
        xmin=10, xmax=1.5e4,
        yticklabels={,,},
        xticklabels={,,},
        enlargelimits=false,
        hide axis
        % before end axis/.code={
        % grid=both,
        % }
    ]
    \addplot[black] table [x=time_steps, y=GPU_par_IEKS, col sep=comma]{runtime.csv};
    \addplot[gray] table [x=time_steps, y=GPU_par_ICKS, col sep=comma]{runtime.csv};
    \addplot +[black, thick, densely dotted, mark=none] coordinates {(3584, \pgfkeysvalueof{/pgfplots/ymin}) (3584, \pgfkeysvalueof{/pgfplots/ymax})};
    \end{axis}
    \end{tikzpicture}%
}}] at (pt){};
% \end{scope}
\end{tikzpicture}}
        \caption{GPU run time}
        \label{fig:gpu}
    \end{subfigure}
    \caption{Run time comparison of the parallel and sequential versions of the IEKS and IPLS on CPU (a) and GPU (b)}
\end{figure}

In Figures \labelcref{fig:cpu,fig:gpu} we observe that while the total computational cost of the parallel implementation of the iterated smoothers is higher than that of their sequential counterparts (Figure \ref{fig:cpu}), the parallelization properties of our proposed algorithms prove beneficial on a distributed environment such as a GPU (Figure \ref{fig:gpu}).
Moreover, as outlined by the medallion in Figure \ref{fig:gpu}, our experiments indeed exhibit the theoretical logarithmic span complexity - derived in \cite{sarkka2020temporal} for a linear Gaussian state space model - up to the parallelization capabilities of our GPU (3584 cores).
\section{conclusion}
\label{sec:conclusion}
In this paper, parallel formulations for two kinds of nonlinear smoothers, namely, iterated sigma-point-based smoothers and iterated extended Kalman smoothers, have been presented. The proposed algorithms have the capability of diminishing the span-complexity from linear to logarithmic. Furthermore, the experimental results, which were conducted on a GPU, showed the benefits of the proposed methods over classical sequential methods.
\vfill\pagebreak
\bibliographystyle{IEEEbib}
\bibliography{strings}
\end{document}